\documentclass{article}
\usepackage[ruled,linesnumbered,vlined]{algorithm2e}
\usepackage{authblk}
\usepackage{amsfonts}
\usepackage{amsmath}
\usepackage{amssymb}
\usepackage{appendix}
\usepackage{array}
\usepackage{caption}
\usepackage{colortbl} 
\usepackage[english]{babel}
\usepackage{eurosym}
\usepackage{graphicx}
\usepackage{bmpsize}
\usepackage{hyperref}
\usepackage{hyphenat}
\usepackage{rotating}
\usepackage{subcaption}
\usepackage{url}
\usepackage{xcolor}
\definecolor{darkgrey}{RGB}{28,28,28} 
\definecolor{mediumgrey}{RGB}{71,71,71} 
\definecolor{lightgrey}{RGB}{115,115,115} 
\definecolor{lightlightgrey}{RGB}{206,206,206} 
\definecolor{darkblue}{RGB}{79,101,140} 
\definecolor{mediumblue}{RGB}{112,144,200} 
\definecolor{lightblue}{RGB}{140,180,250} 
\definecolor{darkgreen}{RGB}{41,115,46} 
\definecolor{mediumgreen}{RGB}{65,180,73} 
\definecolor{lightgreen}{RGB}{90,250,101} 
\definecolor{mediumyellow}{RGB}{247,203,56} 
\definecolor{mediumorange}{RGB}{255,138,60} 
\definecolor{mediumred}{RGB}{219,73,55} 
\definecolor{mediumviolet}{RGB}{146,90,199} 
\definecolor{mediumturquoise}{RGB}{87,189,227} 

\newcommand{\sourcecodeurl}{https://github.com/merschformann/RAWSim-O}

\begin{document}

\title{RAWSim-O: A Simulation Framework for Robotic Mobile Fulfillment Systems}
	
\author[1]{Marius Merschformann\footnote{marius.merschformann@uni-paderborn.de}}
\author[2]{Lin Xie\footnote{lin.xie@leuphana.de}}
\author[3]{Hanyi Li}
\affil[1]{Paderborn University, Paderborn, Germany}
\affil[2]{Leuphana University of L\"uneburg, L\"uneburg, Germany}
\affil[3]{Beijing HANNING ZN Tech Co.,Ltd, Beijing, China}

\date{\today}
\maketitle

\begin{abstract}
This paper deals with a new type of warehousing system, Robotic Mobile Fulfillment Systems (RMFS). In such systems, robots are sent to carry storage units, so-called ``pods,'' from the inventory and bring them to human operators working at stations. At the stations, the items are picked according to customers' orders. There exist new decision problems in such systems, for example, the reallocation of pods after their visits at work stations or the selection of pods to fulfill orders. In order to analyze decision strategies for these decision problems and relations between them, we develop a simulation framework called ``RAWSim-O'' in this paper. Moreover, we show a real-world application of our simulation framework by integrating simple robot prototypes based on vacuum cleaning robots.
\end{abstract}

\section{Introduction}\label{sec:intro}
Due to the rise of e-commerce, the traditional manual picker-to-parts warehousing systems no longer work efficiently, and new types of warehousing systems are required, such as automated parts-to-picker systems. For details about the classification of different warehousing systems we refer to \cite{Koster.2007}. This paper studies one of the parts-to-picker systems, a so-called Robotic Mobile Fulfillment Systems (RMFS), such as the Kiva System (\cite{Enright.2011}, nowadays Amazon Robotics), the GreyOrange Butler or the Swisslog CarryPick. The approach of an RMFS spares some of the time-intense tasks that human workers are confronted with in traditional warehouses, such as the search and retrieval of order items. \cite{Wulfraat.2012} indicates that the Kiva System increases the productivity two to three times, in contrast to a traditional manual picker-to-parts system. Compared to other kinds of warehousing systems, the biggest advantages of an RMFS are its flexibility as a result of having virtually no fixed installations, the scalability due to accessing the inventory in a parallel manner, and the reliability due to the use of only homogeneous components, i.e., redundant components may compensate for faulty ones (see \cite{Hazard.2006} and \cite{Wurman.2008}).

The first framework, ``Alphabet Soup,'' was published by \cite{Hazard.2006}, and is a first simulation of the RMFS concept. In this work, we extend the work of \cite{Hazard.2006} to develop our simulation framework, called ``RAWSim-O'' (Robotic Automatic Warehouse Simulation (for) Optimization). Similarly to ``Alphabet Soup,'' we use an agent-based and event-driven simulation focusing on a detailed view of the system, but extend our simulation framework with the cases of multiple floors, which are connected by elevators. The reason for this extension is to avoid the lack of utilization of vertical space in an RMFS compared to other parts-to-picker systems, such as shuttle-based solutions (see \cite{Tappia.2017}). Moreover, we integrate a more realistic robot movement emulation by considering the robot's turning time and adjusted acceleration or deceleration formulas. This enables a real world application of our simulation framework by using simple robot prototypes based on vacuum cleaning robots (iRobot Create 2). The implementation of the framework was done in C\# and is compatible with the Mono project to allow execution on high throughput clusters. All source codes are available at \url{\sourcecodeurl}.

There exist many decision problems in an RMFS, and they influence each other. With the publication of RAWSim-O we provide a tool for analyzing effects of decision mechanisms for these problems and synergies of decision strategies. The framework enables a holistic look at RMFS which helps uncovering side-effects of strategies, e.g. deciding tasks for robot that cause congestion issues for path planning methods. Hence, RAWSim-O enables a more reliable assessment of the system's overall efficiency, e.g., in terms of customer order throughput. Thus, we hope to further support and push the research on RMFS with our work. We describe the RMFS in more detail and point out the decision problems in focus in Section~\ref{sec:rmfs}. Next, our simulation framework is described in Section~\ref{sec:simulationframework}. After that, we show a demonstration application of our simulation in Section~\ref{sec:demonstration}. Finnaly, we conclude our work in Section~\ref{sec:conclusion}.

\section{The Robotic Mobile Fulfillment System}\label{sec:rmfs}
\begin{figure}[h]
	\centering
	\includegraphics[width=\textwidth]{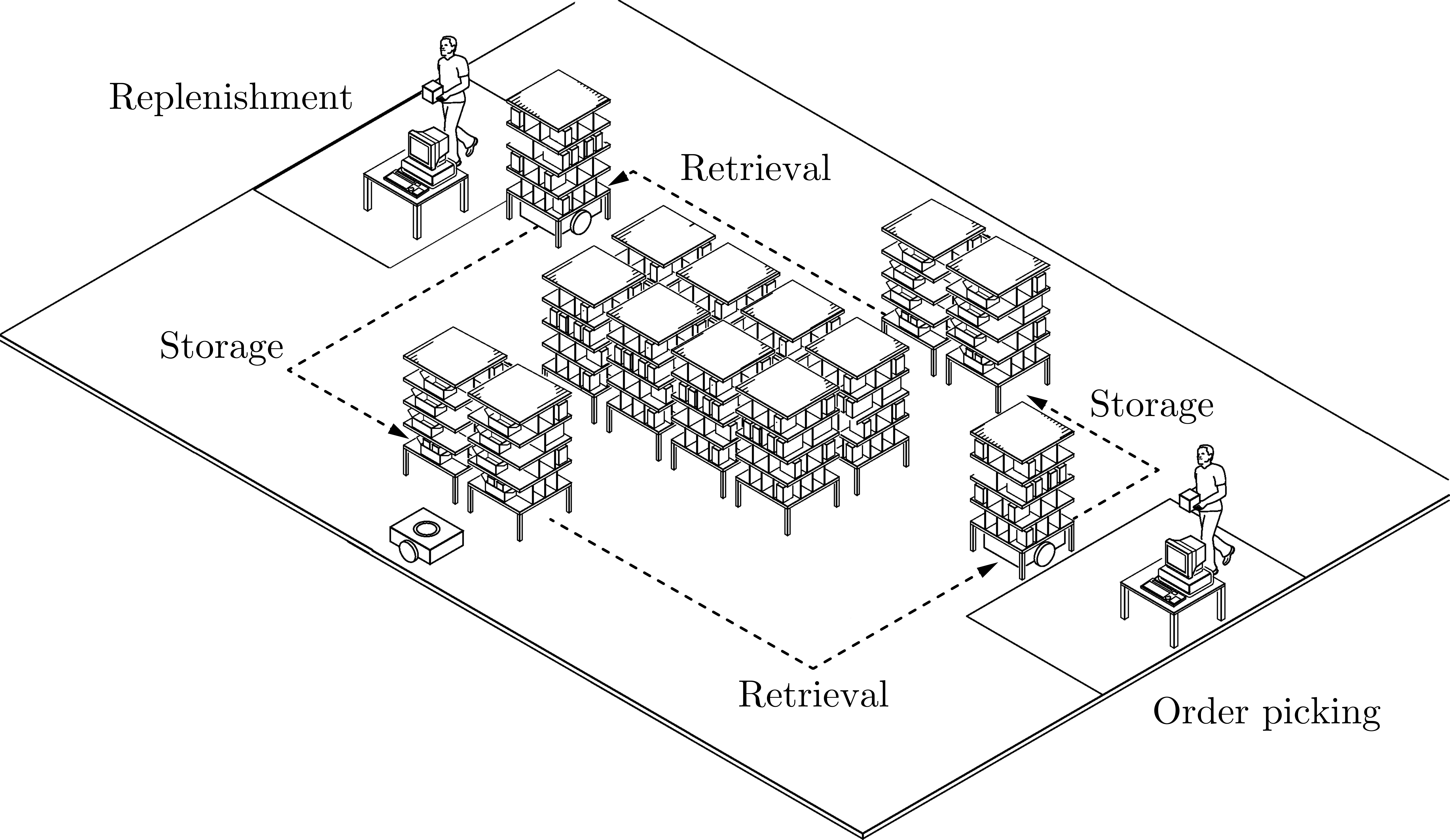}
	\caption{The central process of an RMFS (see \cite{Hoffman.2013})}
	\label{fig:storage_retrieval_process}
\end{figure}

Instead of using a system of shelves and conveyors as in traditional parts-to-picker warehouses,
the central components of an RMFS are:
\begin{itemize}
	\item movable shelves, called \textit{pods}, on which inventory is stored
	\item \textit{storage locations} denoting the inventory area where the pods are stored
	\item workstations, where the pick order items are picked from pods (\textit{pick stations}) or replenishment orders are stored to pods (\textit{replenishment stations})
	\item mobile \textit{robots}, which can move underneath pods and carry them to workstations.
\end{itemize}
The pods are transported by robots between the inventory area and workstations. Figure \ref{fig:storage_retrieval_process} shows the storage and retrieval process: after the arrival of a replenishment order (consisting of a number of physical units of one stock keeping unit (SKU)), robots carry selected pods to a replenishment station to store units in pods. Similarly, after receiving a pick order (including a set of order lines, each for one SKU, with corresponding units necessary to fulfill the line), robots carry selected pods to a pick station, where the units for the order lines are picked. Note that, in order to fulfill pick orders, several pods may be needed, since orders may have multiple lines. Although pods typically contain multiple SKUs with many SKUs in stored in the system it is very unlikely that a pick order can be completed with only one pod. Then, after a pod has been processed at one or more stations, it is brought back to a storage location in the inventory area.

\subsection{Decision Problems}
In an RMFS environment, various optimization and allocation problems have to be solved in real time. The system aims at keeping human workers at the stations busy while minimizing the resources (e.g. robots) to fulfill the incoming pick orders. These problems were first described by \cite{Wurman.2008}:
\begin{itemize} 
	\item For the bots, the planning of their tasks, paths, and motions 
	\item For the stations, the management of the workflow for each station
	\item For the resources of the system, their planning, provisioning, and allocation to other components of the system, usually called the Resource Allocation Problem
\end{itemize}
As \cite{Wurman.2008} note, the Resource Allocation Problem cannot be treated as a global optimization problem, but rather as a set of subproblems that should be solved using specialized methods to enable the decisions in an online e-commerce environment. We reiterate these decision problems briefly in the context of the simulation framework developed in this work:
\begin{itemize}
	\item \textit{Order Assignment}
	\begin{itemize}
		\item \textbf{Replenishment Order Assignment (ROA)}: assignment of replenishment orders to replenishment stations
		\item \textbf{Pick Order Assignment (POA)}: assignment of pick orders to pick stations
	\end{itemize}
	\item \textit{Task Creation}
	\begin{itemize}
		\item \textit{Pod Selection}
		\begin{itemize}
			\item \textbf{Replenishment Pod Selection (RPS)}: selection of the pod to store one replenishment order in
			\item \textbf{Pick Pod Selection (PPS)}: selection of the pods to use for picking the pick orders assigned at a pick station
		\end{itemize}
		\item \textbf{Pod Storage Assignment (PSA)}: assignment of an available storage location to a pod that needs to be brought back to the inventory area
	\end{itemize}
	\item \textbf{Task Allocation (TA)}: assignment of tasks from \textit{Task Creation} and additional support tasks like idling to robots
	\item \textbf{Path Planning (PP)}: planning of the paths for the robots to execute
	\item \textbf{Station Activation (SA)}: controlling active times of the stations (e.g. switching some off based on work shifts or for emulation of ``jumper'' pickers that can be assigned to allow for temporarily increased throughput)
	\item \textbf{Method Management (MM)}: exchange of controlling mechanisms in a running system (e.g. replacing the PSA controller with a different one during runtime in order to adapt to changing conditions)
\end{itemize}
The decisions for the aforementioned operational problems influence each other. Here we sketch two relationships between decision problems that may exploit synergy effects or sabotage each others success:
\begin{itemize}
	\item POA and PPS: one objective is to maximize the
	average number of picks per handled pod (called pile-on); in other words, the higher the pile-on is, the fewer pods are needed at pick stations. Therefore, the selection of a pod for an order in PPS is dependent on the selection of the pick station of other orders. If similar orders are assigned to the same pick station, a pod can be selected to maximize the number of picks.
	\item PSA, TA and PP:  One objective is to minimize the travel times of robots to complete all orders, thus, reducing the number of robots needed to achieve a certain throughput. For this, the selected storage locations for pods impact the performance of the assignment of tasks and the pathfinding, because the trip destinations change.
\end{itemize}

Some of the operational decision problems described above have been addressed in the context of RMFS in previous publications. First, the sequencing of pick orders and pick pods for an integration of PPS and POA is studied in \cite{Boysen.2017}. The authors propose methods for obtaining sequences minimizing the number of pod visits at one station. Second, the planning of paths in an RMFS is subject to the studies in \cite{Cohen.2015}, \cite{Cohen.20170608}, and \cite{Merschformann.2017}. The state-of-the-art multi-agent path planning algorithms were implemented in \cite{Merschformann.2017} to suit PP in an
RMFS. Besides the operational problems, we would like to note that more literature exists on more tactical to strategic decision problems, like layout planning, for example, in \cite{Lamballais.2016}.

\section{RAWSim-O}\label{sec:simulationframework}
RAWSim-O is an agent-based discrete-event simulation framework. It is designed to study the context of an RMFS while considering kinematic behavior of the mobile robots and evaluating multiple decision problems jointly. Hence, the main focus of the framework is the assessment of new control strategies for RMFS and their mutual effects. In the following, we describe the RAWSim-O simulation framework in more detail. At first, we describe the general structure of the framework, and then we give detailed information about how the robotic movement is emulated.

\subsection{Simulation Process}
\begin{figure}[htb]
	\centering
	\begin{subfigure}[b]{0.62\textwidth}
		\includegraphics[width = \textwidth]{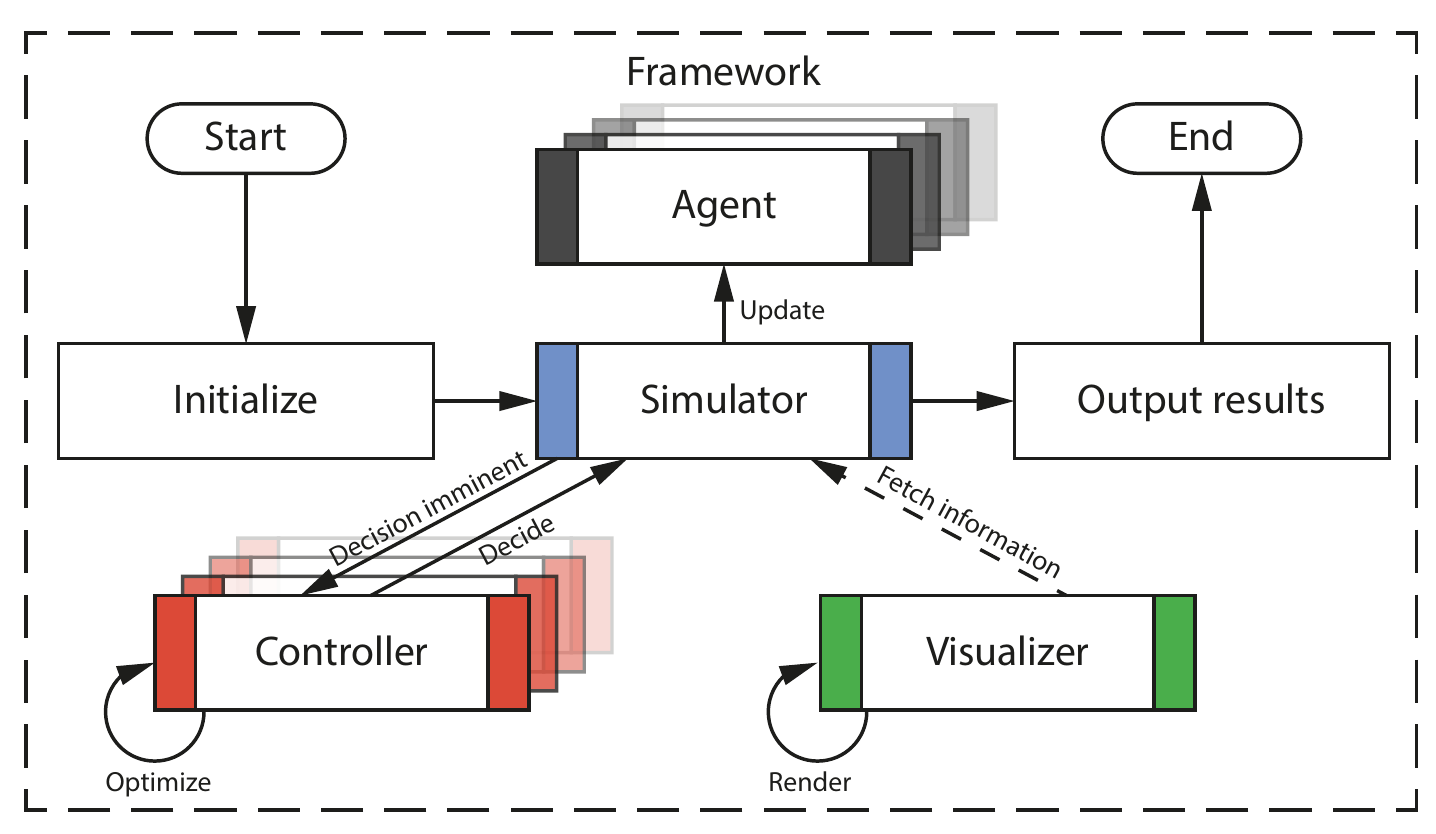}
		\caption{Overview of the simulation process.}
		\label{fig:mu_simulationframework}
	\end{subfigure}
	~
	\begin{subfigure}[b]{0.35\textwidth}
		\includegraphics[width = \textwidth]{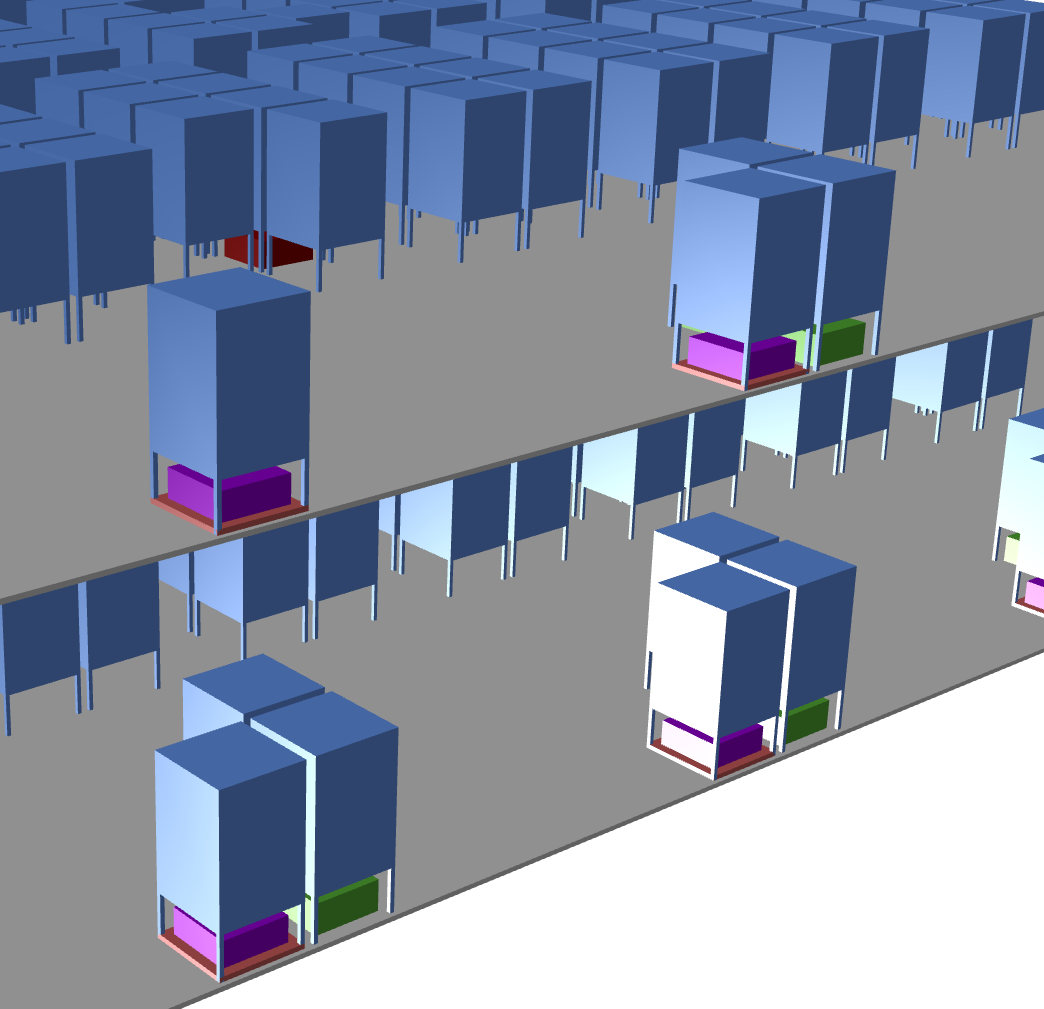}
		\caption{Visualization screenshot}
		\label{fig:mu_simulationscreenshot}
	\end{subfigure}
	\caption{RAWSim-O simulation framework}
\end{figure}

Figure~\ref{fig:mu_simulationframework} shows an overview of our simulation process, which is managed by the core \textit{simulator} instance. The tasks of the simulator contain:
\begin{itemize}
	\item Updating \textit{agents}, which can resemble either real entities, such as robots and stations, or virtual entities like managers, e.g. for emulating order processes.
	\item Passing decisions to \textit{controllers}, which can either  decide immediately or buffer multiple requests and release the decision later. 
	\item Exposing information to a \textit{visualizer}, which allows optional visual feedback in 2D or 3D. Figure~\ref{fig:mu_simulationscreenshot} illustrates a screenshot of our simulation in 3D.
\end{itemize}

As mentioned before, RMFS usually lack utilization of vertical space when compared to other parts-to-picker systems, for example shuttle-based solutions (see \cite{Tappia.2017}). Although the flexibility of the system due to the mobile components is an advantage over the many fixed components of a shuttle-based solution, in areas where land costs are high the lack of vertical space utilization can be a deal-breaker for the RMFS concept. To help mitigate this disadvantage RAWSim-O allows the study of multi-floor systems for applications where the integration of mezzanine floors is possible (see Fig. \ref{fig:mu_simulationscreenshot} for example). The floors are connected by elevators that can be used by one robot at a time. One elevator connects at least two waypoints of the underlying waypoint-system, which is used for guiding the robots. For each pair of connected waypoints a constant time is specified to capture the travel time of the lift transporting a robot.\\
Conceptually, the waypoint system itself is a directed graph which robots travel along while adhering to their kinematic behavior. I.e., turning is only allowed on-the-spot at waypoints (essentially nodes of the graph) and is executed by a specified angular speed. Furthermore, straight travel along the edges of the graph involves acceleration and deceleration, which is discussed in more detail in Sec. \ref{sec:robot_movement_emulation}. To avoid blocked destination waypoints for path planning (many robots share the same destination when approaching stations or elevators) waypoints may form a queuing zone. After a robot reaches such a zone guided by a path planning engine a queue manager takes over handling of the robots paths within the zone. For this, the manager may make use of shortcuts definable within the queue, but mainly lets the robots move up until they reach the end waypoint (e.g., a pick station). On leaving the queuing zone again the defined path planning engine takes over again.

The framework allows easy exchange of controllers by implementing a base structure for all mentioned decision problems. For this, all controllers are also agents of the simulation that may expose a next event to jump to, and also get updated at each event. This enables the mechanisms to react to every change occurring in the simulated world as well as to steer it. As a shortcut, controllers may also subscribe to events that fire, if certain situations occur.\\
Additionally, the controllers are called whenever a new decision needs to be made, e.g. the POA controller is called when a new pick order is submitted to the system, and the TA controller is called when a robot finished its task. In Tab.~\ref{tab:controller-event-overview} we include a short outline of the calls to the event-driven decision controllers. The SA and MM controllers are not event-driven, because they are considered like management mechanisms. However, all controllers may subscribe to simulation-wide events.

\begin{table}[htb]
	\centering
	\caption{Overview of the base events causing calls to the controller per decision problem and the corresponding assignment responsibility}
	\label{tab:controller-event-overview}
\begin{tabular}{l|p{4.4cm}|p{5.5cm}}
Problem & Decision & Main triggers \\
\hline
ROA & repl. order $\rightarrow$ repl. station & New repl. order,\newline repl. order stored in pod \\
POA & pick order $\rightarrow$ pick station & New pick order,\newline pick order completed,\newline repl. order stored in pod \\
RPS & repl. order $\rightarrow$ pod & New repl. order,\newline SKU unit picked \\
PPS & pod $\rightarrow$ pick order(s) & New task is assigned to robot \\
PSA & pod $\rightarrow$ storage location & Pod needs to be brought back to inventory \\
TA & task $\rightarrow$ robot & Robot needs a new task \\
PP & robot $\rightarrow$ robot & Robot has a new destination \\
\end{tabular}
\end{table}

In addition to acting ad-hoc according to the triggers mentioned in Tab.~\ref{tab:controller-event-overview} strategies may plan ahead, e.g., by sequencing tasks for robots instead of greedily selecting them as soon as a robot is done with its previous one. For this example it can be done by preparing tasks ahead and assigning them to the chosen robot as soon as it becomes ready. Furthermore, the framework also allows some controllers to run optimization algorithms in parallel. This means that the controllers are allowed to buffer certain decisions (like the assignment of pick orders), run an optimization procedure in parallel, and submit the decision later on. In this case the simulation is paced until the optimization algorithm returns a solution in order to synchronize the wall time of the algorithm with the simulation time. I.e., the simulation will only continue for the wall time that already passed for the optimization algorithm converted to simulation time (while it is still running), but not beyond it.

\subsubsection{Core Decision Hierarchy}

\begin{figure}[h]
	\centering
	\includegraphics[width=0.85\textwidth]{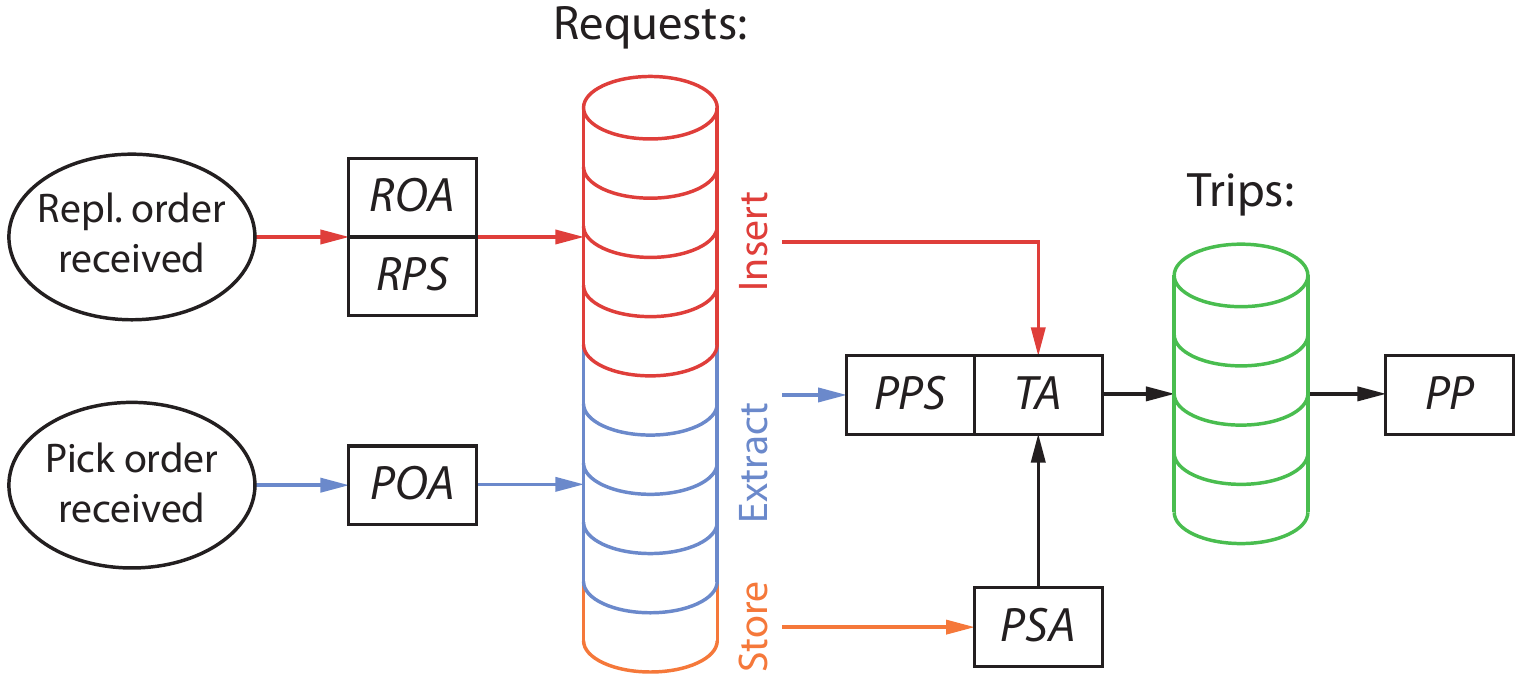}
	\caption{Order of decisions to be done triggered by receiving a pick or replenishment order}
	\label{fig:problemdependencies}
\end{figure}

In the following, we describe the hierarchy of all core decision problems after new replenishment or pick orders are submitted to the system (see Figure~\ref{fig:problemdependencies}). For this, the SA and MM decision problems are neglected, since they have a more supportive role and can even be replaced with default mechanisms that keep the status quo. If a new replenishment order is received, first the controllers of ROA and RPS are responsible for choosing a replenishment station and a pod.
This technically results in an insertion request, i.e. a request for a robot to bring the selected pod to the given workstation.
A number of these requests are then combined in an insertion task and assigned to a robot by a TA controller.
Similarly, after the POA controller selects a pick order from the backlog and assigns it to a pick station, an extraction request is generated, i.e. a request to bring a suitable pod to the chosen station.
Up to this point, the physical units of SKUs for fulfilling the pick order are not yet chosen.
Instead, the decision is postponed and taken just before PPS combines different requests into extraction tasks and TA assigns these tasks to robots.
This allows the implemented controllers to exploit more information when choosing a pod for picking. 
Hence, in this work we consider PPS as a decision closely interlinked with TA. 
Furthermore, the system generates store requests each time a pod is required to be transported to a storage location, and the PSA controller decides the storage location for that pod. The idle robots are located at dwelling points, which are located in the middle of the storage area to avoid blocking prominent storage locations next to the stations. Another possible type of task is charging, if the battery of a robot runs low; however, for this work we assume the battery capacity to be infinite. All of the tasks result in trips, which are planned by a PP algorithm. The only exception is when a pod can be used for another task at the same station, thus, not requiring the robot to move.

\subsubsection{Input Information}

The simulation framework conceptually consists of three different inputs. First, a layout configuration specifies the characteristics and dimensions of the system layout itself. Second, a scenario configuration describes how orders are generated and further settings of the system's surroundings. Finally, a controller configuration is given to specify the decision mechanisms for all previously described decision problems. We distinguish them as three different input files to enable easier assessment of control methods for different systems under diverse scenarios.

\paragraph{Layout Specification}

\begin{figure}[h]
	\centering
	\includegraphics[width=0.9\linewidth]{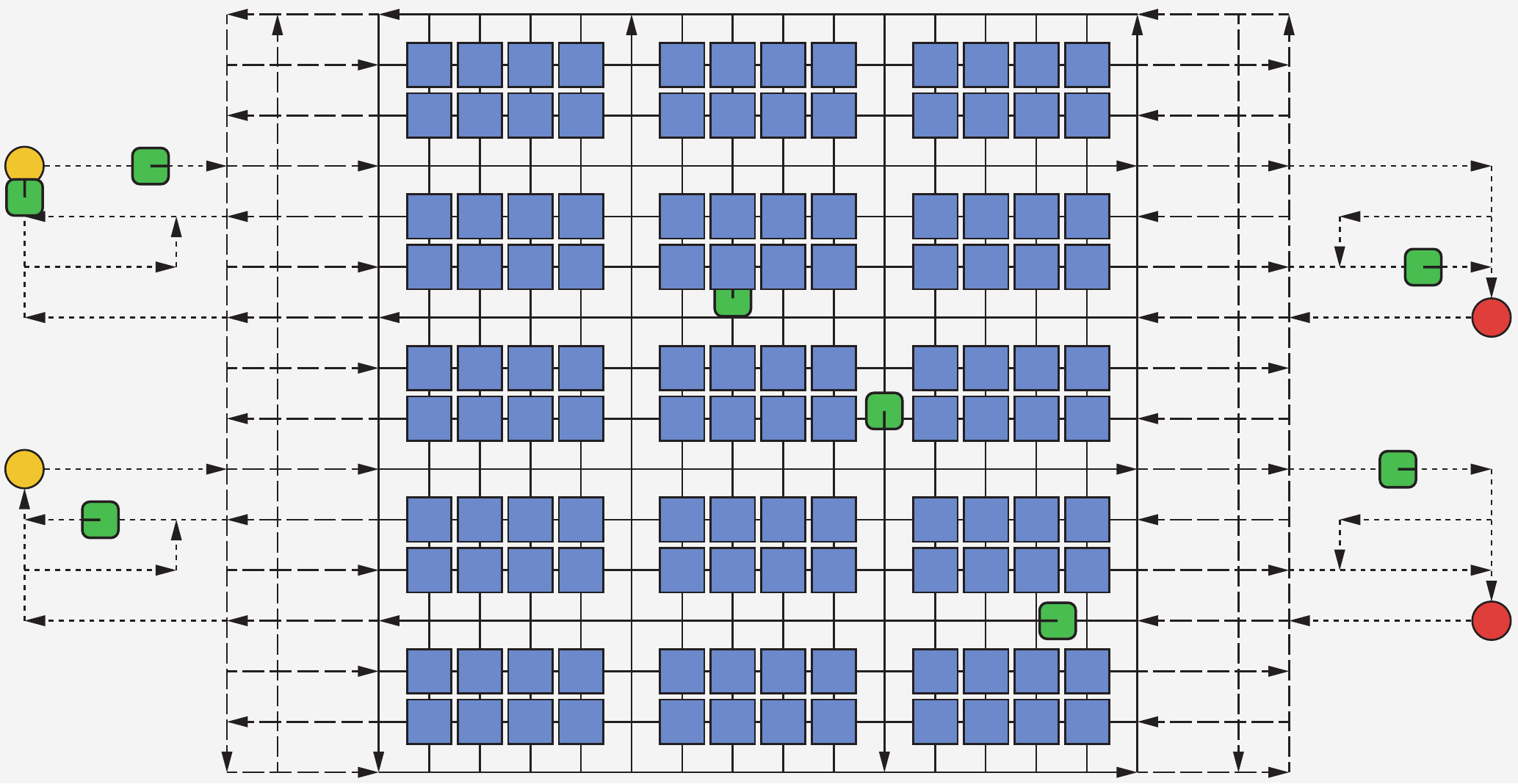}
	\caption{The default layout, with two replenishment and two pick stations}
	\label{fig:mati-layout-2-2-with-bots}
\end{figure}

The layout specification can be either an explicit file specifying the exact positions and individual characteristics for all stations, the waypoint system, and the robots; or a file providing specifications leading to a default layout based on the concepts of \cite{Lamballais.2016}. This default layout can be seen in Fig.~\ref{fig:mati-layout-2-2-with-bots}, with pick stations as red circles, replenishment stations as yellow circles, robots in green, and the pod storage locations in the middle as blue squares. It is based on the idea of circular flows around storage location blocks that also align with the entrances and exits of the stations and their queuing areas (dotted lines). Between the queuing area and the storage locations, a hallway area (dashed lines) allows the robots to cross between the stations. In order to generate such a layout, the numbers of pick and replenishment stations on the north, south, east, and west sides need to be set. Furthermore, the numbers of vertical and horizontal aisles and the block size need to be specified.

\paragraph{Scenario Specification}
The next input file specifies information about the scenario to simulate. This includes the duration of the simulation and settings about the inventory and order emulation. In contrast with the concept of colored word order emulation of ``Alphabet Soup'' in \cite{Hazard.2006}, we use a concept for SKU and pick order generation based on typical random distributions. For the generation of SKU popularity information, we implemented constant, uniform, normal, and gamma distributions to emulate simple scenarios up to more ABC-like popularity curves. The SKU for each pick order line is then selected using the chosen distribution, but limiting the choice to only in-stock products. This is done to avoid stock-out situations, which are not useful when simulating the control of an RMFS for such situation, i.e. we assume that no unfulfillable pick orders are submitted to the system. The number of order lines and the number of units per line are also chosen from a distribution previously specified by the user. For choosing the SKU for each replenishment order, we use the same popularity distribution connected to information about the order size in which a certain product is replenished. For this, we also allow setting a parameterized amount of replenishment orders to be return orders, i.e. single line and single unit replenishment orders. The space consumption of one unit of a SKU on a pod is a one-dimensional factor which is also set according to a user-specified distribution. This is necessary to enable the simulation of both pick and replenishment operations at the same time, while emulating the inventory situation in our system.

Moreover, we implement three procedures for generating new replenishment and pick orders. First, a constant order backlog scenario can be selected. This means that a completed order is immediately replaced by a new one. This is done to keep the system under constant pressure. To avoid completely overfilling or draining the inventory over time, we allow the specification of replenishment and pick order generation pauses according to given inventory level thresholds. E.g., if the inventory reaches a 95\% filling, replenishment order generation is paused until it drops below 50\% again. Second, we generate orders according to arrivals of a configurable Poisson process. This Poisson process can be inhomogeneous to capture order peak situations during a day or match certain patterns observed at distribution centers. Finally, we allow the input of files to specify which orders are generated during the simulation horizon. Additionally, we allow a combined setting of one of the order generation scenarios above with the submission of order batches at given periodic time points. This is done to allow emulation of batch operations or hybrid scenarios. With all of these options, we aim to resemble most artificial and realistic scenarios.

\paragraph{Controller Configuration}

The last input file specifies the controller to use for each decision problem as well as the parameters. This enables a flexible configuration due to the modular controller concept. Controllers integrating multiple decision problems can be configured using dummy controllers for the other components. With the help of reflection most parameter structures only need to be defined once. Without additional implementation effort new parameters are serializable and configurable in the graphical user interface.

\begin{figure}[htb]
	\centering
	\begin{subfigure}[b]{0.57\textwidth}
		\includegraphics[width = \textwidth]{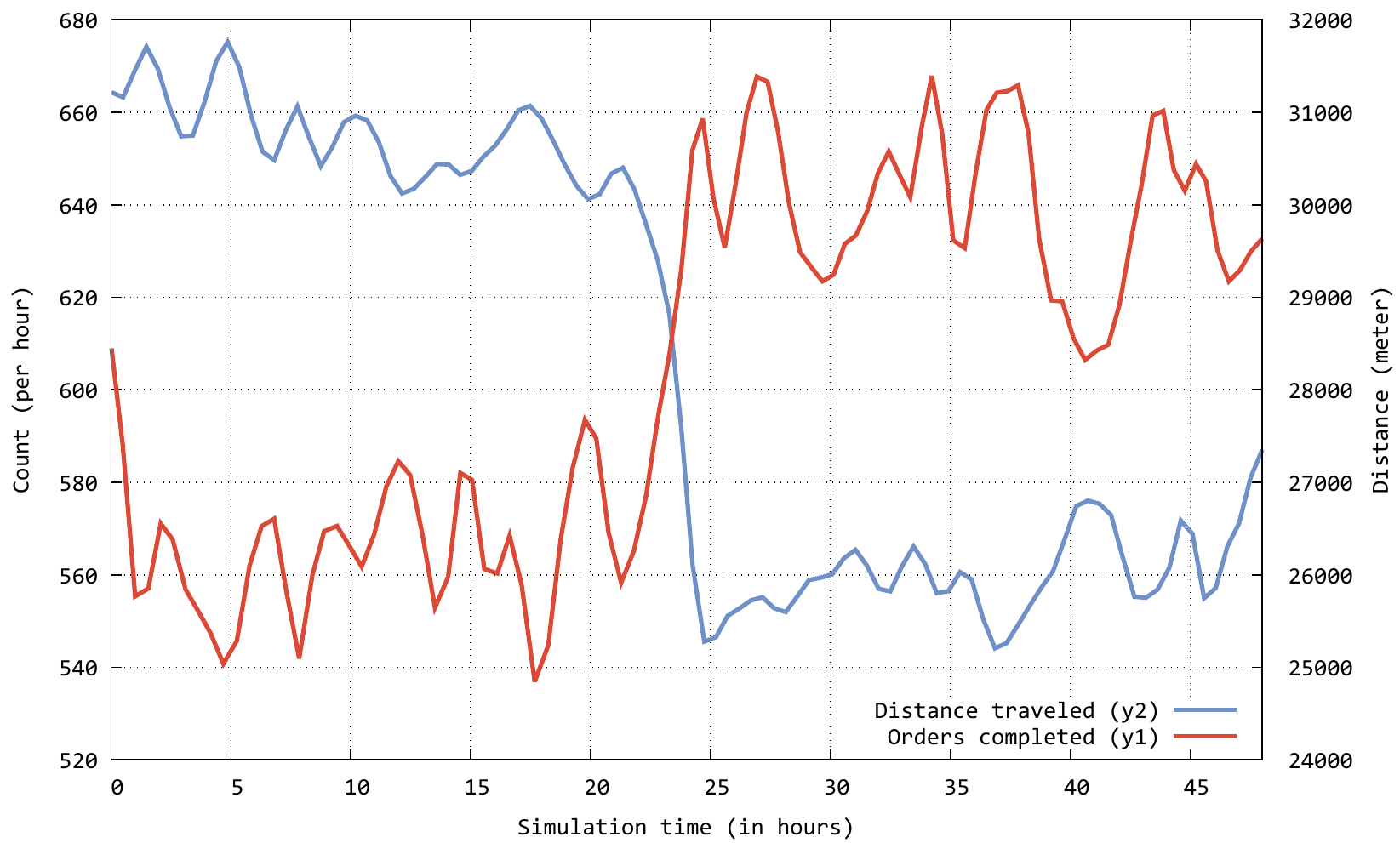}
		\caption{Sample progression plot of picked orders (red) and distance traveled (blue) per hour}
		\label{fig:plot_example}
	\end{subfigure}
~
\begin{subfigure}[b]{0.4\textwidth}
	\includegraphics[width = \textwidth]{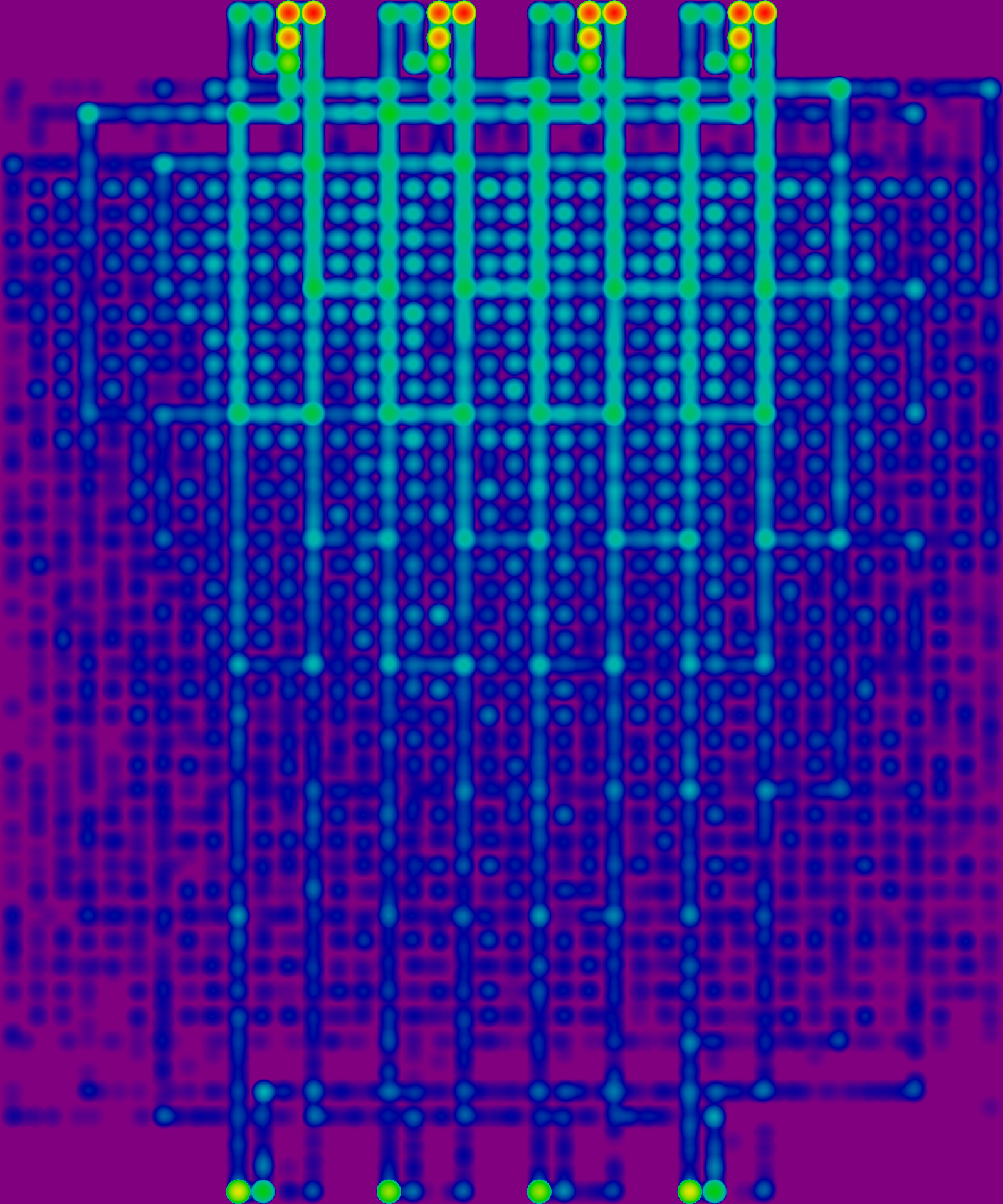}
	\caption{Example of a heatmap showing robot movement behavior over time (red $\equiv$ high, purple $\equiv$ low)}
	\label{fig:heatmap_example}
\end{subfigure}
\caption{Examples of the system's output}
\end{figure}

\subsubsection{Output}

In order to investigate the system's behavior in more detail, different output measures are tracked and logged over time. For this, we distinguish three main measure types. First, a footprint is written for each simulation execution to allow for an easy comparison of multiple executions. This footprint contains most basic performance measures for the simulation run like the order throughput rate or the distance traveled by the robots. Second, time-based information is logged to generate plots after execution of the simulation. These are useful for post-experiment analysis of the simulation processes, especially if no visualization was attached. As an example, Figure~\ref{fig:plot_example} shows the pick order throughput and overall distance traveled per hour for a simulation in which the PSA strategy is switched after half of the simulation horizon. For this experiment, a random PSA strategy was replaced with a strategy selecting the nearest available storage location for the pod. This is reflected by the decrease in distance to cover by the robots, which immediately leads to an increased throughput in pick orders, because more pods are available at the pick stations. Finally, location-based information is logged to conduct heatmap analyses. This is useful, for example, to get insights about congestion effects when looking at robot movement behavior over time (see Fig.~\ref{fig:heatmap_example}). In this example, we can see the queuing happening at the pick stations at the top and the replenishment stations at the bottom. Furthermore, we can identify highly frequented areas which are prone to congestion. Note that the logarithm was applied to the heat values in order to increase the contrast in color in the resulting heatmap.

\subsection{Robot Movement Emulation}\label{sec:robot_movement_emulation}
\newcommand{\mIndexTime}{t}
\newcommand{\mPhysicsTime}{t}
\newcommand{\mPhysicsPosition}[1]{s_{#1}}
\newcommand{\mPhysicsDistance}{d}
\newcommand{\mPhysicsSpeed}[1]{v_{#1}}
\newcommand{\mPhysicsTopSpeed}{\overline{v}}
\newcommand{\mPhysicsAcceleration}{\overrightarrow{a}}
\newcommand{\mPhysicsDeceleration}{\overleftarrow{a}}
\newcommand{\mPhysicsTimeUntilTopSpeed}{\mPhysicsTime_{\mPhysicsSpeed{0} \rightarrow \mPhysicsTopSpeed}}
\newcommand{\mPhysicsDistanceUntilTopSpeed}{\mPhysicsPosition{\mPhysicsTimeUntilTopSpeed}}
\newcommand{\mPhysicsTimeUntilFullDeceleration}{\mPhysicsTime_{\mPhysicsSpeed{0} \rightarrow 0}}
\newcommand{\mPhysicsDistanceUntilFullDeceleration}{\mPhysicsPosition{\mPhysicsTimeUntilFullDeceleration}}

\begin{table}[thb]
	\caption{Symbol definitions}
	\centering
	\begin{tabular}{l|l}\label{tab:physics-symbols}
		Symbol & Description \\
		\hline
		\ $\mPhysicsSpeed{\mIndexTime}$ & The speed at time $\mIndexTime$ \\
		\ $\mPhysicsPosition{\mIndexTime}$ & The position at time $\mIndexTime$ \\
		\ $\mPhysicsAcceleration$ & Acceleration in $\frac{m}{s^2}$ \\
		\ $\mPhysicsDeceleration$ & Deceleration in $\frac{m}{s^2}$ (negative) \\
		\ $\mPhysicsTopSpeed$ & Top-speed in $\frac{m}{s}$ \\
	\end{tabular}
\end{table}

The calculations of traveling times and distances for straight movement are based on uniform acceleration and deceleration. For this, the velocity of a robot has to be considered to determine its arrival time at a destination node. The symbols used in the following description are defined in Tab.~\ref{tab:physics-symbols}. The fundamental formulas for all remaining definitions are given by the speed (see Eq.~\ref{eq:physics-basic-speed}), the difference in position while accelerating (see Eq.~\ref{eq:physics-basic-positionacceleration}), and the difference in position during top-speed (see Eq.~\ref{eq:physics-basic-positiontopspeed}).
\begin{align}
	\label{eq:physics-basic-speed}
	\mPhysicsSpeed{\mPhysicsTime} &= \mPhysicsAcceleration \mPhysicsTime + \mPhysicsSpeed{0} \\
	\label{eq:physics-basic-positionacceleration}
	\mPhysicsPosition{\mPhysicsTime} &= \frac{\mPhysicsAcceleration}{2} \mPhysicsTime^2 + \mPhysicsSpeed{0} \mPhysicsTime + \mPhysicsPosition{0} \\
	\label{eq:physics-basic-positiontopspeed}
	\mPhysicsPosition{\mPhysicsTime} &= \mPhysicsTopSpeed \mPhysicsTime + \mPhysicsPosition{0}
\end{align}
For the implementation of the simulation framework, it is required to not only calculate the time for covering a distance when initially standing still, but also to calculate distances and times based on an initial speed $v_0 > 0$, because simulation events may occur at any time during robot travel. For this, the time $\left( \mPhysicsTimeUntilTopSpeed \right)$ and distance $\left( \mPhysicsDistanceUntilTopSpeed \right)$ to reach the top-speed are needed (see Eq.~\ref{eq:physics-acc-timeuntiltopspeed}~\&~\ref{eq:physics-acc-distanceuntiltopspeed}). This can analogously be defined for full deceleration (see Eq.~\ref{eq:physics-dec-timeuntiltopspeed}~\&~\ref{eq:physics-dec-distanceuntiltopspeed}).
\begin{align}
	\label{eq:physics-acc-timeuntiltopspeed}
	\mPhysicsTopSpeed &= \mPhysicsAcceleration \mPhysicsTimeUntilTopSpeed + \mPhysicsSpeed{0} \Leftrightarrow \mPhysicsTimeUntilTopSpeed = \frac{\mPhysicsTopSpeed - \mPhysicsSpeed{0}}{\mPhysicsAcceleration}\\
	\label{eq:physics-acc-distanceuntiltopspeed}
	\mPhysicsDistanceUntilTopSpeed &= \frac{\mPhysicsAcceleration}{2} \left( \mPhysicsTimeUntilTopSpeed \right)^2 + \mPhysicsSpeed{0} \mPhysicsTimeUntilTopSpeed\\
	\label{eq:physics-dec-timeuntiltopspeed}
	0 &= \mPhysicsDeceleration \mPhysicsTimeUntilFullDeceleration + \mPhysicsSpeed{0} \Leftrightarrow \mPhysicsTimeUntilFullDeceleration = \frac{0 - \mPhysicsSpeed{0}}{\mPhysicsDeceleration}\\
	\label{eq:physics-dec-distanceuntiltopspeed}
	\mPhysicsDistanceUntilFullDeceleration &= \frac{\mPhysicsDeceleration}{2} \left( \mPhysicsTimeUntilFullDeceleration \right)^2 + \mPhysicsSpeed{0} \mPhysicsTimeUntilFullDeceleration
\end{align}
For the calculation of time for traveling distance $\mPhysicsDistance$ when starting at an initial speed of $\mPhysicsSpeed{0}$, four cases need to be considered. In the first case, only deceleration is possible. In the second case, cruising at top-speed and deceleration are possible. In the third case, acceleration up to top-speed, cruising at top-speed and deceleration are possible. In the fourth case, the distance is so short that only acceleration and deceleration phases are possible. The function defined in Alg.~\ref{alg:physics-calculate-time} calculates the remaining cruise time for all of the cases. Line~\ref{alg:physics-calculate-time-acc-and-dec-only} uses the time at which acceleration switches to deceleration, which is described in more detail below.

\begin{algorithm}[tbh]
	\caption{$CruiseTime(\protect\mPhysicsAcceleration, \protect\mPhysicsDeceleration, \mPhysicsTopSpeed, \mPhysicsSpeed{0}, \mPhysicsDistance)$}
	\label{alg:physics-calculate-time}
	\DontPrintSemicolon
	\LinesNumbered
	\If{$\mPhysicsDistance \le \mPhysicsDistanceUntilFullDeceleration$}{
		\Return{$\mPhysicsTimeUntilFullDeceleration$}
	}
	\If{$\mPhysicsSpeed{0} = \mPhysicsTopSpeed$}{
		\Return{$\frac{\mPhysicsDistance - \mPhysicsDistanceUntilFullDeceleration}{\mPhysicsTopSpeed} + \mPhysicsTimeUntilFullDeceleration$}
	}
	\If{$\mPhysicsDistanceUntilFullDeceleration + \mPhysicsDistanceUntilFullDeceleration \le \mPhysicsDistance$}{
		\Return{$\mPhysicsTimeUntilTopSpeed + \frac{\mPhysicsDistance - \mPhysicsDistanceUntilTopSpeed - \mPhysicsDistanceUntilFullDeceleration}{\mPhysicsTopSpeed} + \mPhysicsTimeUntilFullDeceleration$}
	}
	\Return{$\sqrt{\dfrac{\mPhysicsDistance + \frac{\mPhysicsAcceleration}{2}\left(\frac{\mPhysicsSpeed{0}}{\mPhysicsAcceleration}\right)^2}{\frac{\mPhysicsAcceleration}{2}+\frac{\mPhysicsAcceleration}{2\mPhysicsDeceleration}}} + \sqrt{\dfrac{\mPhysicsDistance + \frac{\mPhysicsAcceleration}{2}\left(\frac{\mPhysicsSpeed{0}}{\mPhysicsAcceleration}\right)^2}{\frac{\mPhysicsDeceleration}{2}+\frac{\mPhysicsDeceleration}{2\mPhysicsAcceleration}}} - \dfrac{\mPhysicsSpeed{0}}{\mPhysicsAcceleration}$}\label{alg:physics-calculate-time-acc-and-dec-only}
\end{algorithm}

Let $\mPhysicsDistance'$ be the full distance from the start node to the destination node, thus, starting with zero speed at the beginning of $\mPhysicsDistance'$ and stopping with zero speed at the end of $\mPhysicsDistance'$. For this, the time of switching from acceleration to deceleration is given by Eq.~\ref{eq:physics-acc-dec-switch-time-first}, i.e., the time for driving while accelerating. To calculate this time we make use of the fact that the speed at which we switch from acceleration to deceleration must match ($\mPhysicsAcceleration \mPhysicsTime_1 = \mPhysicsDeceleration \mPhysicsTime_2$), hence, we can substitute $\mPhysicsTime_2$ with $\frac{\mPhysicsAcceleration \mPhysicsTime_1}{\mPhysicsDeceleration}$.
\begin{align}\label{eq:physics-acc-dec-switch-time-first}
	\mPhysicsDistance' &= \frac{\mPhysicsAcceleration}{2} \mPhysicsTime^2_1 + \frac{\mPhysicsDeceleration}{2} \mPhysicsTime^2_2 \nonumber \\
	\Leftrightarrow \mPhysicsDistance' &= \frac{\mPhysicsAcceleration}{2} \mPhysicsTime^2_1 + \frac{\mPhysicsDeceleration}{2} \left( \frac{\mPhysicsAcceleration \mPhysicsTime_1}{\mPhysicsDeceleration} \right)^2 \nonumber \\
	\Leftrightarrow \mPhysicsDistance' &= \frac{\mPhysicsAcceleration}{2} \mPhysicsTime^2_1 + \frac{\mPhysicsDeceleration}{2} \frac{\left( \mPhysicsAcceleration \mPhysicsTime_1 \right)^2}{2\mPhysicsDeceleration} \nonumber \\
	\Leftrightarrow \mPhysicsTime_1^2 &= \frac{\mPhysicsDistance'}{\frac{\mPhysicsAcceleration}{2} + \frac{\mPhysicsAcceleration^2}{2\mPhysicsDeceleration}} \nonumber \\
	\Leftrightarrow \mPhysicsTime_1 &= \sqrt{\frac{\mPhysicsDistance'}{\frac{\mPhysicsAcceleration}{2} + \frac{\mPhysicsAcceleration^2}{2\mPhysicsDeceleration}}}
\end{align}
For the calculation, we assume that speed is currently zero and the movement just starts at the start node. If $\mPhysicsSpeed{0} > 0$ and $\mPhysicsDistance < \mPhysicsDistance'$, then $\mPhysicsDistance'$ has to be calculated by using $\mPhysicsDistance$ (see Eq.~\ref{eq:physics-acc-dec-switch-time-second}).
\begin{equation}\label{eq:physics-acc-dec-switch-time-second}
\mPhysicsDistance' = \mPhysicsDistance + \frac{\mPhysicsAcceleration}{2} \left( \frac{\mPhysicsSpeed{0}}{\mPhysicsAcceleration} \right)^2
\end{equation}
Analogously to Eq.~\ref{eq:physics-acc-dec-switch-time-first}, it is also possible to solve for $\mPhysicsTime_2$, i.e., the time for driving while decelerating. The sum of $\mPhysicsTime_1$ and $\mPhysicsTime_2$ is the complete time for the cruise from start node to destination node. The time for accelerating to $\mPhysicsSpeed{0}$ is being subtracted, such that the remaining time can be expressed as in line~\ref{alg:physics-calculate-time-acc-and-dec-only} of Alg.~\ref{alg:physics-calculate-time}.

\section{Demonstrator}\label{sec:demonstration}
\begin{figure}[htb]
	\centering
	\begin{subfigure}[b]{0.52\textwidth}
		\includegraphics[width=\textwidth]{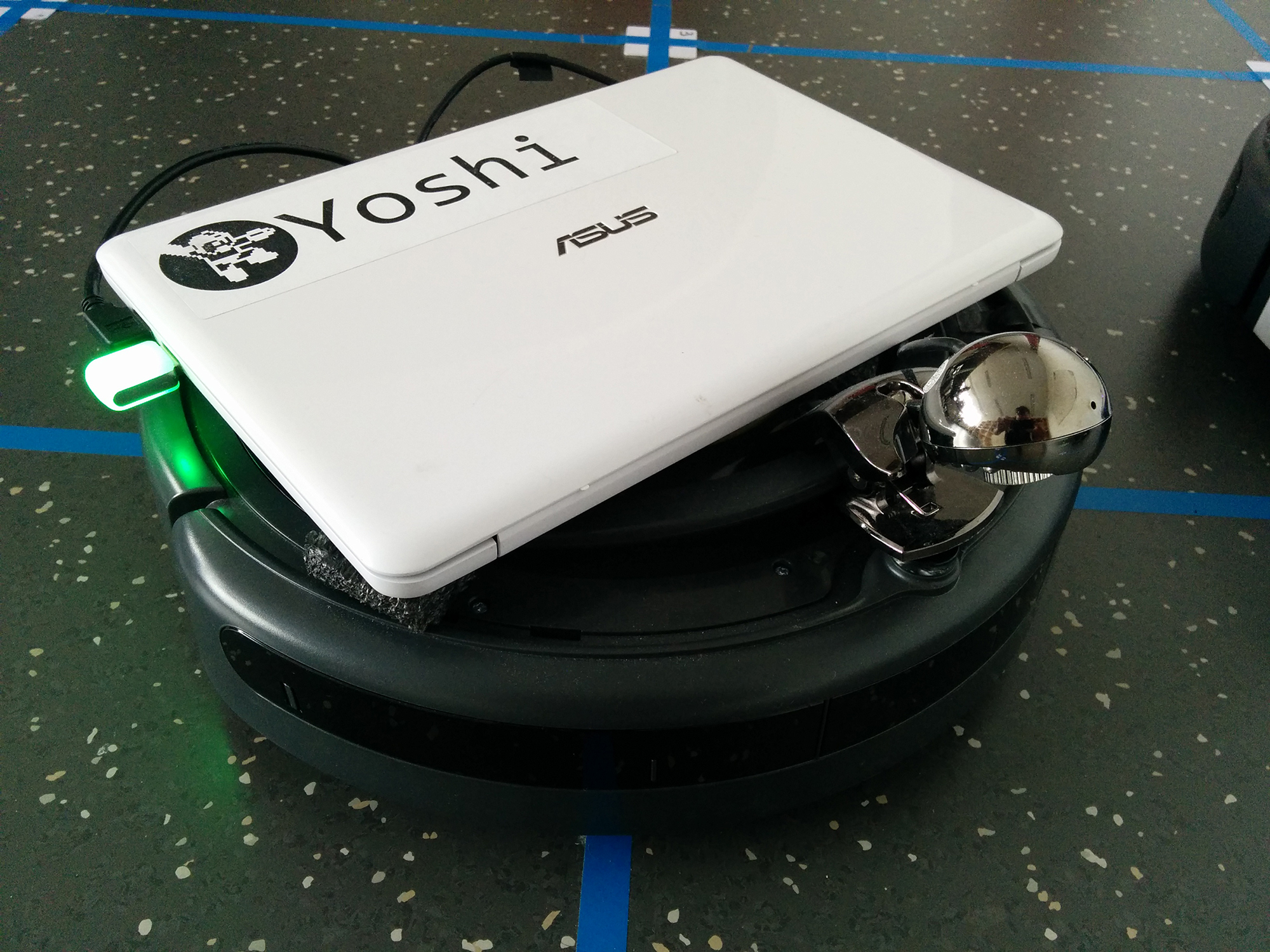}
		\caption{Close-up of a demonstrator robot}
		\label{fig:demonstrator_onerobot}
	\end{subfigure}
	~
	\begin{subfigure}[b]{0.42\textwidth}
		\includegraphics[width=\textwidth]{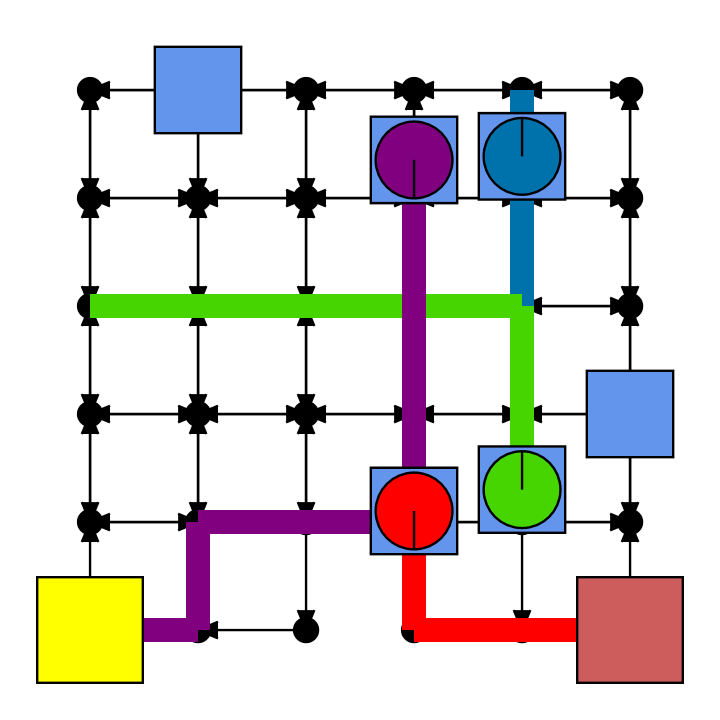}
		\caption{The view from RAWSim-O}
		\label{fig:simulator_multiplerobots}
	\end{subfigure}
	~
	\begin{subfigure}[b]{0.52\textwidth}
		\includegraphics[width=\textwidth]{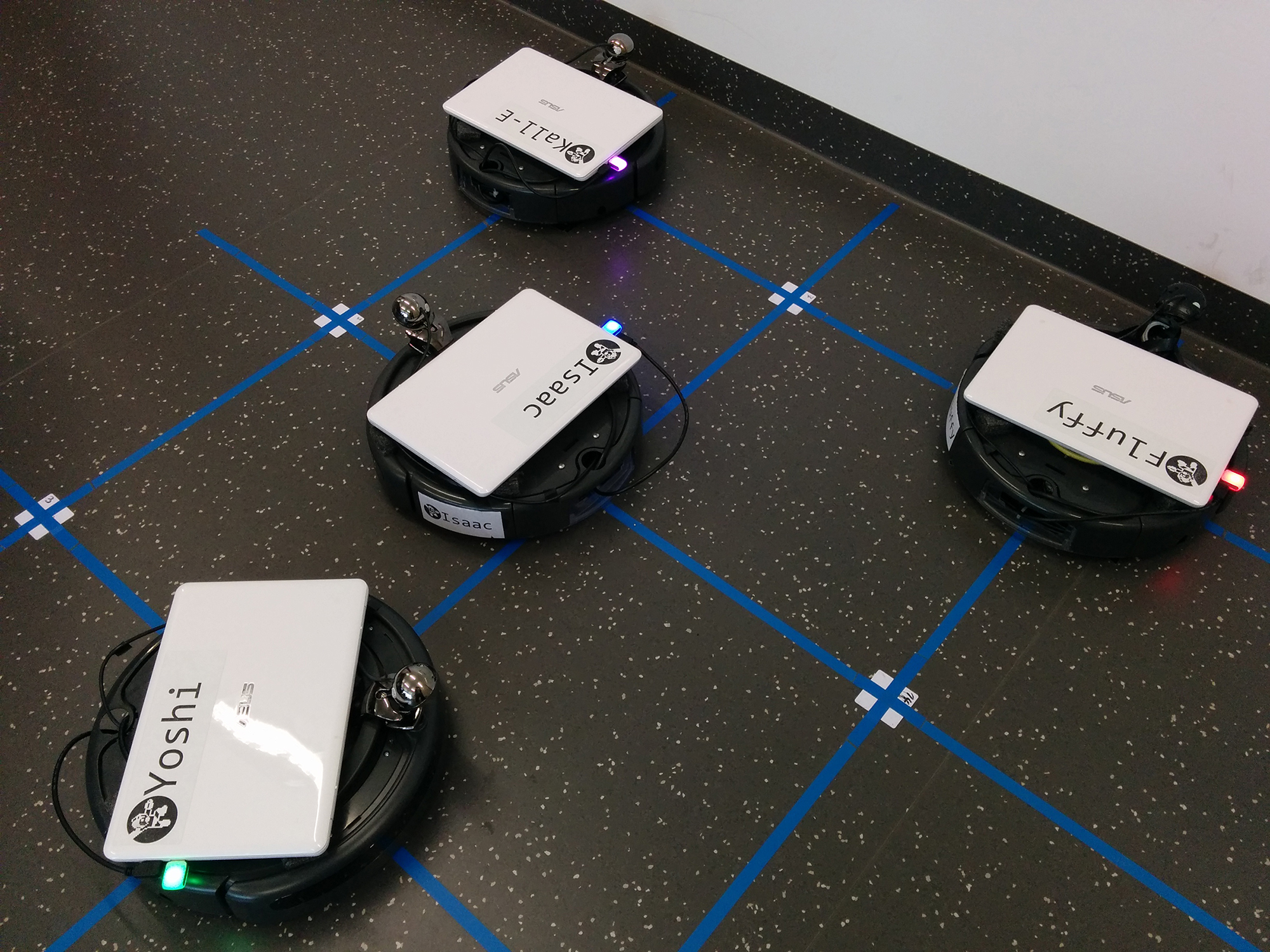}
		\caption{Multiple robots in an emulated RMFS}
		\label{fig:demonstrator_multiplerobots}
	\end{subfigure}
	\caption{Demonstrator example of four robots running in an emulated RMFS}
\end{figure}

We implemented a demonstrator functionality to investigate how our algorithms developed within RAWSim-O work with real robots. For example with the help of the iRobot Create 2, a mobile robot platform based on the Roomba vacuum cleaning robot. There are several reasons we choose the iRobot Create 2: ease of programming, low complexity, similar movement behavior like typical RMFS robots and low costs. The robots are equipped with ASUS Eee PCs through serial-to-USB cables for processing capabilities, webcams for line-following, blink(1) for visual feedback, and RFID tag readers mounted inside the former vacuum cleaning compartment for waypoint recognition (see Figure~\ref{fig:demonstrator_onerobot}). Although we cannot emulate the transportation of pods with the robots, we are still able to study the overall movement behavior in a real situation, which is more prone to errors and noise than a simulation. Technically, the demonstrator robots replace the simulated robots, hence, a hybrid of a real system and a simulation is built.

We demonstrate our RMFS with a simple running example of four robots in a grid-world with $0.45 m \times 0.45 m$ cells (totally 36 cells), see Figure~\ref{fig:simulator_multiplerobots} for the view from RAWSim-O for this example and Figure~\ref{fig:demonstrator_multiplerobots} for the view of the demonstration. One replenishment station is set on the left bottom side and one pick station is set on the right bottom side. In total, there are six pods (blue rectangles) available. The maximum velocity limit of each robot is $0.21 m/s$, while the time it takes for each robot to do a complete turn is set to $5.5s$. And the maximum acceleration and deceleration of each robot are set to $0.5$ and $-0.5$. To adhere to the kinematic constraints (such as turning times and acceleration) in continious time, we use the MAPFWR-solver from \cite{Merschformann.2017} to generate time-efficient collision- and deadlock-free paths. Those paths are converted to a sequence of go straight, turn left, and turn right commands and sent to the robot via WiFi. The robot then executes these commands and sends back the RFID waypoint tag of each intersection it comes across. By doing this, the movement of the real robot and the expected movement are synchronized. This needs to be done in order to avoid errors from the noisy real-world setting to add up. Similarly to the simulated environment, pick and replenishment operations are emulated by blocking the robot at the station for a fixed time. With the published source code and similar components the demonstrator can be rebuild.

\begin{figure}[htb]
	\centering
	\begin{subfigure}[b]{0.48\textwidth}
		\includegraphics[width=\textwidth]{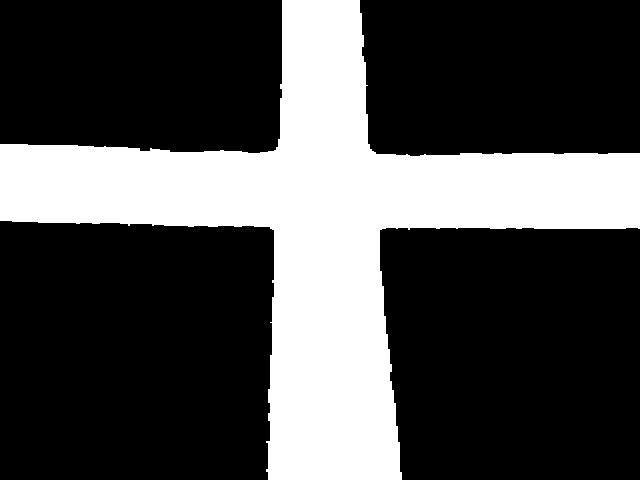}
		\caption{Filtered image showing the line}
		\label{fig:linefollower_bw}
	\end{subfigure}
	~
	\begin{subfigure}[b]{0.48\textwidth}
		\includegraphics[width=\textwidth]{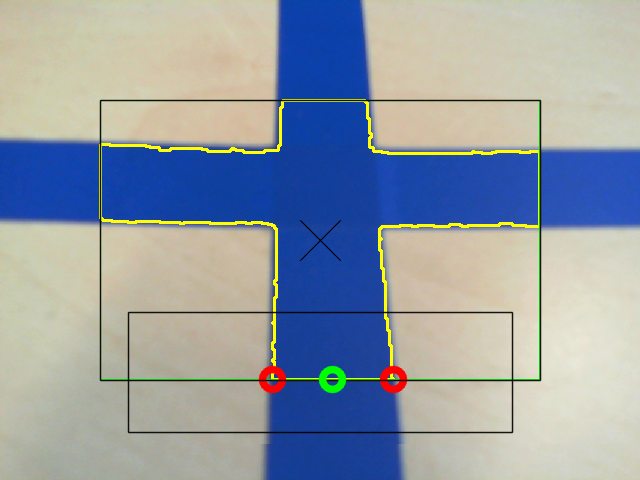}
		\caption{Processing result on original image}
		\label{fig:linefollower_color}
	\end{subfigure}
	\caption{Line-following example using a blue line}
\end{figure}

The robot line-following is implemented using a common PID (Proportional Integration Differential) controller method. The center of the line is extracted by using the OpenCV library. The line recognition first translates the image to a binary one by applying a range filter, and then applies the basic morphological operation ``erosion'' for removing noise (see resulting image in Figure \ref{fig:linefollower_bw}). Afterwards, contour extraction is used to find the line's borders within the gray image. Using these lines, the bottom center is estimated to allow a stable target (see green ring in Figure \ref{fig:linefollower_color}) for the PID controller.\\
Using the demonstrator, we are able to show the successful application of the path planning algorithms developed within RAWSim-O in a real-world situation. Furthermore, it can be used to demonstrate the basic idea of an RMFS at a small scale.

\section{Conclusion}\label{sec:conclusion}
In this work, we outlined core real-time decision problems occurring when operating an RMFS. To investigate different solution approaches for those decision problems, we introduce the RMFS simulation framework RAWSim-O and describe its functionality and capabilities. Alongside this publication, we also publish the source code of the framework and already present controllers and algorithms at \url{\sourcecodeurl} to support future research on RMFS. With this work we aim to support future research on RMFS. For example, a study of how to control systems involving multiple mezzanine floors can be done using RAWSim-O. For these, an efficient storage strategy and task allocation method is expected to be crucial in order to mitigate the bottle-neck effect introduced by the elevators. Furthermore, we have shown demonstrator application capabilities of the framework that enable the integration of simple robots for real world demonstrations.

\section{Acknowledgements}
We would like to thank Tim Lamballais for providing us with the concepts and implementation of the default layout generator used in RAWSim-O.

\bibliographystyle{plain}
\bibliography{ms}

\end{document}